\documentclass[11pt]{article}
\usepackage{times}
\usepackage{sidecap}
\usepackage{geometry}
\usepackage{floatrow}
\usepackage{float}
\usepackage{hyperref}
\usepackage{graphics,graphicx}
\usepackage{subcaption}
\usepackage{epsf,wrapfig}
\usepackage{titlesec}

\usepackage{natbib,bm}

\usepackage{paralist}
\let\olditem\item
\renewenvironment{thebibliography}[1]{%
  \section*{\refname}
  \let\par\relax
  \renewcommand{\item}[1][]{\olditem[\textbullet]}%
  \inparaenum}{\endinparaenum}

\newcounter{aff}

\DeclareGraphicsExtensions{.jpg,.pdf,.png,.eps}

\usepackage{caption}
\usepackage{xcolor}

\usepackage{tcolorbox}

\bibpunct{(}{)}{;}{a}{}{,}

\usepackage{soul}

\definecolor{darkcyan}{rgb}{0.0, 0.0, 0.7}

\definecolor{boxcolor}{RGB}{193, 229, 255}
\definecolor{figboxcolor}{RGB}{193, 229, 255}
\definecolor{boxcolorgray}{RGB}{227, 227, 227}
\definecolor{darkcyan}{rgb}{0.0, 0.0, 0.7}

\sethlcolor{boxcolorgray}

\def\gtrsim{\lower 2pt \hbox{$\, \buildrel {\scriptstyle >}\over
{\scriptstyle \sim}\,$}}
\def\lesssim{\lower 2pt \hbox{$\, \buildrel {\scriptstyle <}\over
{\scriptstyle \sim}\,$}}

\def\approxlt{\lower.2em\hbox{$\buildrel < \over \sim$}}
\def\approxgt{\lower.2em\hbox{$\buildrel > \over \sim$}}

\def\be{\begin{equation}}
\def\ee{\end{equation}}
\def\bea{\begin{eqnarray*}}
\def\eea{\end{eqnarray*}}

\def\jref@jnl#1{{\rm#1}}
\def\aj{\jref@jnl{AJ}}
\def\araa{\jref@jnl{ARA\&A}}
\def\apj{\jref@jnl{ApJ}}
\def\apjl{\jref@jnl{ApJ}}
\def\apjs{\jref@jnl{ApJS}}
\def\ao{\jref@jnl{Appl.~Opt.}}
\def\apss{\jref@jnl{Ap\&SS}}
\def\aap{\jref@jnl{A\&A}}
\def\aapr{\jref@jnl{A\&A~Rev.}}
\def\aaps{\jref@jnl{A\&AS}}
\def\azh{\jref@jnl{AZh}}
\def\baas{\jref@jnl{BAAS}}
\def\jrasc{\jref@jnl{JRASC}}
\def\memras{\jref@jnl{MmRAS}}
\def\mnras{\jref@jnl{MNRAS}}
\def\pra{\jref@jnl{Phys.~Rev.~A}}
\def\prb{\jref@jnl{Phys.~Rev.~B}}
\def\prc{\jref@jnl{Phys.~Rev.~C}}
\def\prd{\jref@jnl{Phys.~Rev.~D}}
\def\pre{\jref@jnl{Phys.~Rev.~E}}
\def\prl{\jref@jnl{Phys.~Rev.~Lett.}}
\def\psj{\jref@jnl{PSJ}}
\def\pasp{\jref@jnl{PASP}}
\def\pasa{\jref@jnl{PASA}}
\def\pasj{\jref@jnl{PASJ}}
\def\qjras{\jref@jnl{QJRAS}}
\def\skytel{\jref@jnl{S\&T}}
\def\solphys{\jref@jnl{Sol.~Phys.}}
\def\sovast{\jref@jnl{Soviet~Ast.}}
\def\ssr{\jref@jnl{Space~Sci.~Rev.}}
\def\zap{\jref@jnl{ZAp}}
\def\nat{\jref@jnl{Nature}}
\def\iaucirc{\jref@jnl{IAU~Circ.}}
\def\aplett{\jref@jnl{Astrophys.~Lett.}}
\def\apspr{\jref@jnl{Astrophys.~Space~Phys.~Res.}}
\def\bain{\jref@jnl{Bull.~Astron.~Inst.~Netherlands}}
\def\fcp{\jref@jnl{Fund.~Cosmic~Phys.}}
\def\gca{\jref@jnl{Geochim.~Cosmochim.~Acta}}
\def\grl{\jref@jnl{Geophys.~Res.~Lett.}}
\def\jcp{\jref@jnl{J.~Chem.~Phys.}}
\def\jgr{\jref@jnl{J.~Geophys.~Res.}}
\def\jqsrt{\jref@jnl{J.~Quant.~Spec.~Radiat.~Transf.}}
\def\memsai{\jref@jnl{Mem.~Soc.~Astron.~Italiana}}
\def\nphysa{\jref@jnl{Nucl.~Phys.~A}}
\def\physrep{\jref@jnl{Phys.~Rep.}}
\def\physscr{\jref@jnl{Phys.~Scr}}
\def\planss{\jref@jnl{Planet.~Space~Sci.}}
\def\procspie{\jref@jnl{Proc.~SPIE}}

\def\ion#1#2{#1$\;${\small\rm\@Roman{#2}}\relax}

\newcount\lecurrentfam
\def\LaTeX{\lecurrentfam=\the\fam \leavevmode L\raise.42ex
\hbox{$\fam\lecurrentfam\scriptstyle\kern-.3em A$}\kern-.15em\TeX}
\def\sizrpt{
(\fontname\the\font): em=\the\fontdimen6\font, ex=\the\fontdimen5\font
\typeout{
(\fontname\the\font): em=\the\fontdimen6\font, ex=\the\fontdimen5\font
}}
\definecolor{dark}{rgb}{ 0.34, 0.09, 0.49}

\usepackage{fancyhdr,url}
\usepackage{transparent}
\usepackage{tablefootnote}
\usepackage{mathpazo}

\titlespacing*{\paragraph}{0pt}{1.0ex plus 0.5ex minus 0.2ex}{1em}

\begin{document}

\noindent
\begin{raggedright}
{\Huge\bfseries Evolved stars, multiplicity and planets  \\[0.5em]
\Large Shaping and Being Shaped by Stellar Evolution
}\end{raggedright}\\

\thispagestyle{empty}
\noindent
\vspace{-0.7cm}
\begin{flushleft}

\noindent
Theo Khouri$^{1}$, Wouter Vlemmings$^{1}$,
Elvire De Beck$^{1}$, Tomek Kami\'nski$^2$, Taïssa Danilovich$^{3,4}$, Marie Van de Sande$^5$,
Devika Kamath$^6$, Maksym Mohorian$^6$, Shazrene Mohamed$^{7,8}$, Alain Baudry$^9$, Raghvendra Sahai$^{10}$, Carmen S\'anchez-Contreras$^{11}$, Javier Alcolea$^{12}$, Luis Velilla-Prieto$^{13}$, L\'ea Planquart$^1$, Leen Decin$^{14}$, Pierre Kervella$^{15,16}$, Bannawit Pimpanuwat$^{17}$, Matthias Maercker$^{1}$, Sandra Etoka$^{18}$, Behzad Bojnordi Arbab$^1$, Ramlal Unnikrishnan$^8$, Guillermo Quintana-Lacaci$^{13}$, Miguel Santander-Garc\'ia$^{12}$
\vspace{1em}
{
\footnotesize{
\noindent
\\
$^{1}$Department of Physics and Astronomy, Chalmers University of Technology, SE-412 96 Gothenburg, Sweden\\
$^2$Nicolaus Copernicus Astronomical Center, Polish Academy of Sciences\\
$^3$University of Newcastle, Australia\\
$^4$Anton Pannekoek Institute for Astronomy, University of Amsterdam, Netherlands\\
$^5$School of Physics, University College Dublin, Belfield, Dublin 4, Ireland\\
$^6$School of Mathematical and Physical Sciences, Macquarie University, Balaclava Road, Sydney, NSW 2109, Australia\\
$^7$Department of Astronomy, University of Virginia, 530 McCormick Road, Charlottesville, VA 22904, USA\\
$8$Virginia Institute for Theoretical Astronomy, University of Virginia, Charlottesville, VA 22904, USA
$^9$Univ. of Bordeaux, L.A.B., France\\
$^{10}$Jet Propulsion Laboratory, California Institute of Technology, Pasadena, CA 91109, USA\\
$^{11}$Centro de Astrobiologia (CAB), CSIC/INTA, Spain\\
$^{12}$Observatorio Astron\'omico Nacional, IGN/GNIG, MITMS, Spain\\
$^{13}$Institute of Fundamental Physics (IFF), CSIC, Spain\\
$^{14}$KU Leuven, Belgium\\
$^{15}$LIRA, Observatoire de Paris, Université PSL, Sorbonne Université, Université Paris Cité, CY Cergy Paris Université
$^{16}$French-Chilean Laboratory for Astronomy, IRL 3386, CNRS and U. de Chile, Casilla 36-D, Santiago, Chile\\
$^{17}$National Astronomical Research Institute of Thailand (Public Organization), Chiang Mai 50180, Thailand\\
$^{18}$Jodrell Bank Centre for Astrophysics, The University of Manchester, Manchester M13 9PL, UK\\
}}
\footnotetext[1]{\url{theo.khouri@chalmers.se}}

\begin{tcolorbox}[boxsep=3pt,left=0pt,right=0pt,top=1pt,bottom=1pt,colframe=boxcolor,colback=boxcolor]
\justify
\textbf{Abstract:}
Understanding how planets and stellar companions survive and interact with their host stars as they evolve, or are ultimately destroyed in the process, is essential both for predicting the long-term future of planetary systems like our own and the contribution of stars to galactic chemical evolution.
In this white paper, we discuss how the next generation ALMA (ALMA2040) will revolutionize our understanding of the key processes by detecting and characterising accreting companions and planetary engulfment signatures in hundreds of systems with evolved stars.
\end{tcolorbox}
\end{flushleft}

\vspace{-0.7cm}
\section{Scientific context and open questions}

Our understanding of critical phases of stellar evolution has advanced significantly in recent years with improvements in our ability to empirically characterize and theoretically simulate key processes such as mass loss \citep{Khouri2018,Sahai2022,VelillaPrieto2023,Khouri2024,Freytag2023,ahmad2023}, convection \citep{freytag2017, paladini2018_pi1gru,vlemmings2024_rdor}, and chemical complexity \citep{Agundez2020,Van-de-Sande2022,Danilovich2024,Wallstrom2024,Alonso-Hernandez2025}.
These followed from the development of sensitive, high-resolution instruments such as ALMA, VLT, and VLTI, and
advancements in computational power and techniques that allowed the simulation of complex interactions of micro and macro physics to an unprecedented degree.

One of the most consequential open questions in the field of stellar evolution is how the presence of companions affects the fates of stars. This has also been prominently highlighted by observations of evolved systems since the advent of ALMA.
In fact, one of the important empirical results of the last decade is that the effects of stellar and/or planetary companions on the wind of evolved stars are ubiquitous, even for evolved stars not originally expected to be experiencing or to have experienced interactions with a companion \citep{Maercker2012,Vlemmings2018,Decin2020}.
These effects can be spectacular, through mergers and interactions that eject the whole stellar envelope \citep[e.g.][]{Ivanova2013} and produce complex morphologies and jets \citep[e.g.][]{Sahai2007,Tafoya2020}, or less spectacular but very consequential through enhancing mass loss \citep{Decin2021, Aydi2022}. The progeny of systems that interacted spans a large number of classes and phenomena: sub-dwarf B stars, Barium stars, blue stragglers, ellipsoidal variables, novae, luminous red novae, systems with white dwarf (WD) companions, double-degenerate systems, many planetary nebulae, Type Ia Supernovae, cataclysmic variables, symbiotic binaries, X-ray binaries, and gravitational wave sources. This non-exhaustive list spans interactions occurring at every stage of stellar evolution, from the main sequence to the remnant phase. 
The processes controlling binary interactions have far-reaching consequences. As discussed below, our understanding is still severely lacking. ALMA2040 will allow avenues of investigation completely impossible at the moment, opening up the way for the advancement of our empirical understanding. In the paragraphs below, we describe the types of processes and astrophysical objects that are relevant in this context. In Section~\ref{sec:sciencecase}, we describe objective ways in which ALMA2040 will contribute.

\paragraph{Stellar multiplicity
and interactions:}
When stars evolve to become
red (super)giants, their sizes increase dramatically. This potentially leads to mass transfer, tidal interactions, and common-envelope evolution that reshape the configuration of a binary or multiple system. Although mass loss, tidal torques, and angular momentum redistribution are tightly coupled, we lack empirical characterization of how these processes drive orbital evolution beyond the main sequence.
Whereas astrometric missions such as Hipparcos and Gaia and (optical) radial velocity studies allow for the characterization of earlier-type red giant branch (RGB) systems and more evolved post-asymptotic giant branch (post-AGB) stars, only a few AGB and red supergiant (RSG) systems have well-characterised orbital configurations. While this is expected to dramatically change in the near future with upcoming Gaia data releases, the direct study of the interaction processes will remain scarce.

One clear point of tension is the distributions of orbital distance and eccentricity observed in systems that experienced substantial mass transfer. Effectively all samples of binary interaction products, such as Ba stars, post-AGB binaries, WD binaries, CEMP-s stars, and extrinsic S stars, do not match the predictions of current mass-transfer and tidal theory \citep{Abate2015,Mathieu2025,vanSon2026}. Possible solutions include better description of the mass-transfer processes that are more stable than currently considered \citep[e.g.][]{Woods2011, Ge2020, Temmink2023}, higher accretion efficiency \citep{Mohamed2007},
 formation of circumbinary discs \citep{Chen2020} and
orbital shrinking from angular momentum loss \citep[e.g.][]{Jahanara2005,Saladino2019}. 
From an empirical perspective, it is clear that interaction physics is poorly constrained. ALMA2040 will completely change this landscape by allowing studies of the key processes in samples of objects in the evolutionary phase where the interaction is the strongest and the most consequential, when one of the components swells to giant dimensions. 

Mass transfer from an evolved star to a companion can occur via three distinct regimes depending on how the Roche lobe radius compares to its stellar and dust condensation radii. At the widest separations the wind reaches terminal velocity before encountering the companion and accretion follows the Bondi-Hoyle-Lyttleton prescription \citep{Hoyle1939,Bondi1944}, with efficiencies (accreted mass relative to mass lost)
of order a few percent. If the dust 
condensation radius is comparable to or larger than the Roche lobe radius, the wind is gravitationally focused by the companion, leading to what is known as Wind Roche Lobe Overflow \citep[WRLOF,][]{Mohamed2007,Mohamed2012}. Simulations suggest WRLOF significantly increases the accretion efficiency from a few percent up to tens of percent of the mass lost \citep{Mohamed2012,Abate2013}. At still smaller separations, the  photosphere itself fills the Roche lobe. Whether such systems survive as 
persistent mass transfer binaries or enter common envelope depends on the poorly understood stability of mass transfer from convective giant donors.

\begin{figure}
  \includegraphics[width=0.35\textwidth,trim={10cm 6cm 10cm 6cm},clip]{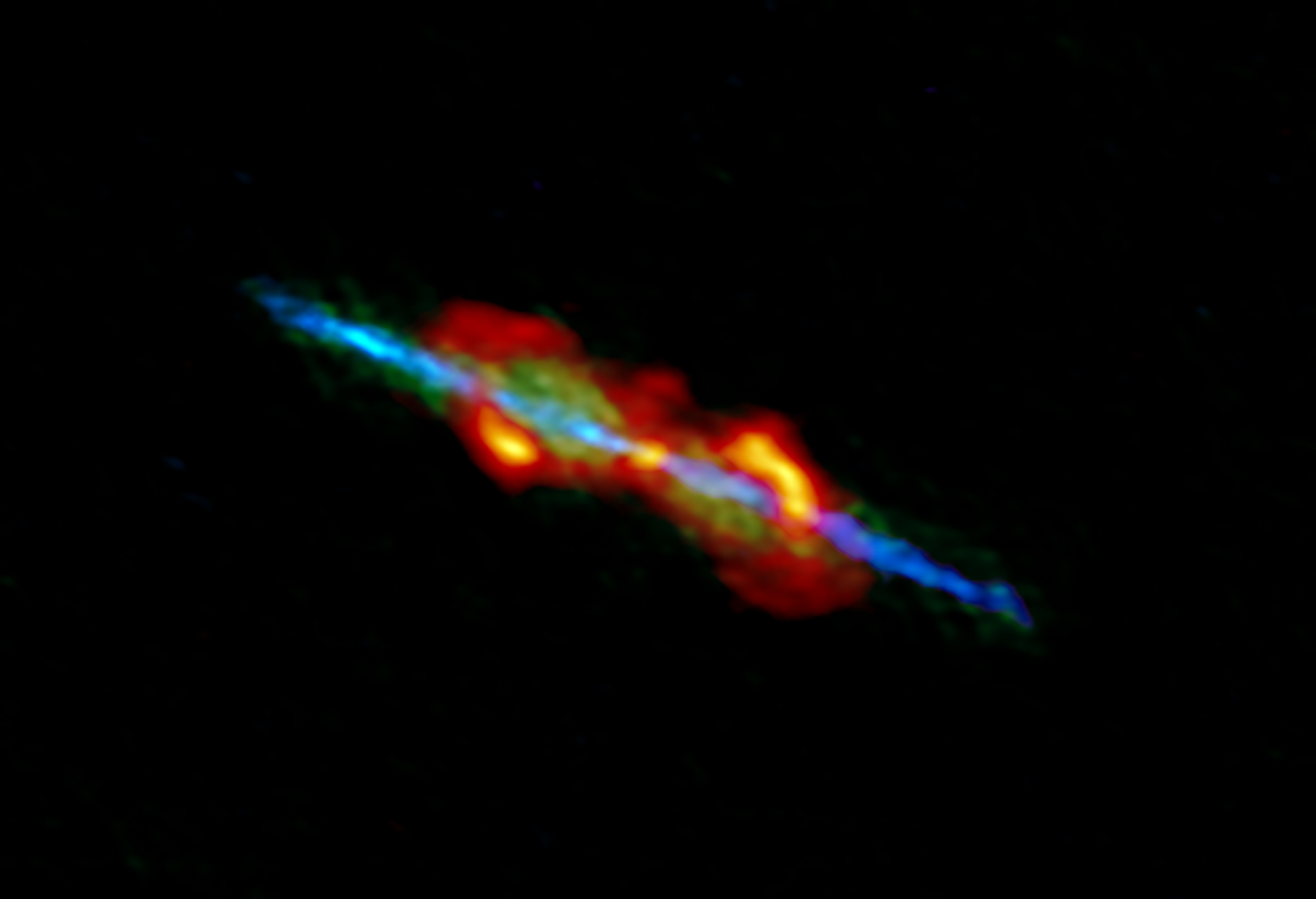} 
\includegraphics[width=0.25\textwidth]{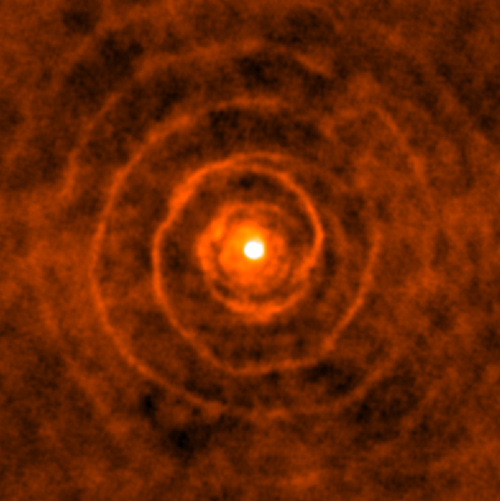} 
 \includegraphics[width=0.35\linewidth,trim={7cm 3.4cm 12cm 0cm},clip]{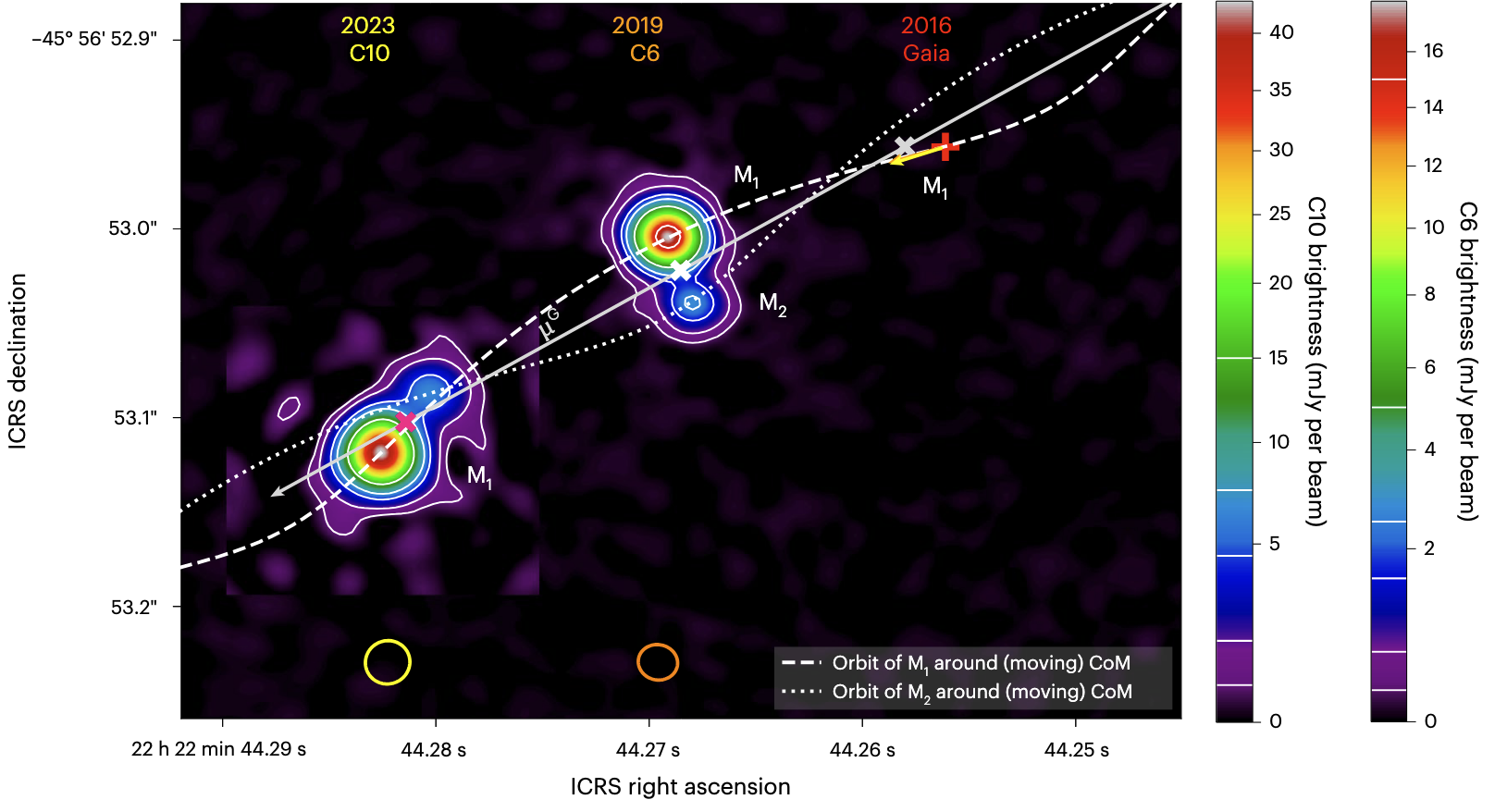}
\caption{Examples of ALMA observations of systems where interactions between an evolved star and a companion can be studied. \textit{Left:} the water fountain star W43~A suspected to be a post-common-envelope object. The image shows dust emission (heat map) and high-velocity ($100 > v > 75$~km/s, blue) and low-velocity ($v < 50$~km/s, green) CO emission \citep{Tafoya2020}.
\textit{Middle:} the clear spiral pattern imprinted in the outflow from the carbon-rich AGB star AFGL~3068 by the presence of a companion on an eccentric orbit \citep{kim2017_afgl3068}.
\textit{Right:} The orbital motion of a companion star, likely on the Main Sequence, around the AGB star $\pi^1$~Gru, measured by ALMA and further constrained by Hipparcos and Gaia observations \citep{esseldeurs2025_pigru}.}
\label{fig:examples}
\end{figure}

Unfortunately, advancements in empirical characterization of these processes are hindered by the fact
that very few systems are known to be in a phase of strong interaction (when one of the components is at giant dimensions ), because
of the intrinsic difficulty in finding low-mass main sequence stars close to bright giants, e.g. within $\lesssim 15$~milliarcseconds (mas) for separation $\lesssim 45$~au  at $\sim 300$~pc. As shown in Section~\ref{sec:sciencecase}, roughly ten accreting companions to an evolved star are expected to be detectable in total with the current capabilities of ALMA. This explains why only a small number of main sequence companions to evolved stars in compact orbits are suspected or known \citep[e.g. $\pi^1$~Gru and V~Hya,][right panel of Fig.~\ref{fig:examples}]{Homan2020,Planquart2024,esseldeurs2025_pigru}.
Greater sensitivity and angular resolution would raise this number to hundreds, opening a new era in the study of mass transfer and orbital evolution in systems with giant stars. Additionally, the characterization of orbits using the astrometry of the two stars (also using data from Gaia) will allow for the determination of the system mass, providing key information that is rarely available in the study of AGB stars. In this way, ALMA2040 would provide an overview of accretion in systems spanning a wide range of orbital configurations.

\paragraph{Post-interaction systems and transients:}
The end products of strong binary interactions -- dusty (post-)AGB binaries (Fig.~\ref{fig:examples}, left panel), merger remnants, and red transients -- are among the most complex and least understood phases of stellar evolution. Many post-interaction systems are known or suspected \citep[e.g.,][]{BoffinJones,Kaminski2021}, with several classes of evolved stars now recognized or suggested to be outcomes of recent binary interaction, such as post-AGB stars and extreme AGB stars \citep[e.g.][]{Jones2015,DellAgli2021,Khouri2021,Khouri2025,VanWinckel2025}, newly identified post-RGB objects \citep[e.g.][]{Kamath2015,Sahai,olofsson2019,Moltzer2025}, and red or blue stragglers \citep[e.g.][]{patrick2020}.
Common-envelope and merger events produce luminous red novae transients, but these remain sparsely characterized \citep{Karambelkar2025}. Due to a large amount of circumstellar dust,
many of these systems lack optical counterparts and can be studied only at infrared and (sub-)millimeter wavelengths \citep[e.g.][]{Kami2018,steinmetz}.
The interaction processes that control the evolution of these systems remain poorly constrained.

A fraction of the systems that interact during the transition from the AGB (RGB) to the post-AGB (post-RGB) phase form circumbinary disks \citep{VanWinckel2003}. These disks act as reservoirs of mass and angular momentum and continue to interact with the central system: re-accretion feeds discs and jets around the companion
\citep{Oomen2020,Mohorian2024,VanWinckel2025}, while the binary in turn reshapes the disk and drives disk winds \citep{Verhamme2024}. 
These disks have cavity edges, spirals, warps, and wind-launching regions on scales of a few to tens of au (angular sizes of a few mas for typical distance of $\sim 2$~kpc). The details of the disk structure relate directly to orbital evolution \citep{Kluska2022}.

An exciting question is whether these disks are potential birth-places for second-generation giant planets. Although gravitational fragmentation appears unlikely under present-day post-AGB disk conditions \citep{Pourmand2025}, planets could form through core accretion. Such planets could be detected in the future through compact continuum emission from circumplanetary material, or indirectly through gaps, dust traps, azimuthal asymmetries, and departures from Keplerian rotation — the same diagnostics that reveal planet--disk interaction in protoplanetary disks \citep{Andrews2018,Zhang2018}. Distinguishing binary-driven structures from companion-carved cavities requires resolving both dust morphology and gas kinematics on au scales \citep{Corporaal2023}. IRAS~08544$-$4431 is an immediate benchmark: at $\sim 1.6$~kpc, current 10~mas observations probe 
$\sim 16$~au and barely resolve its inner rim, modelled at 16.5~au in diameter \citep[Mohorian et al. accepted;][]{Kluska2018}.
ALMA2040 will also be crucial to identify observables which can differentiate between disk-planet interactions with potential planets from the first or second generation.

\paragraph{Planets:}
The last decades have seen enormous advances regarding our understanding of the properties of planetary systems, with more than 6500 exoplanets detected to date \citep[NASA Exoplanet Archive;][]{NASAexoplanetarchive2025} and far more expected by the 2040s. However, almost all known systems orbit main-sequence hosts. The post-main-sequence evolution of these systems -- whether planets survive engulfment, are ejected, or accrete new material \citep[e.g.,][]{Veras,Mustill,Spiegel} -- remains largely unexplored and at most a few substellar companions of evolved stars will be known by 2040. The next generation facility ALMA2040 will completely reshape our understanding by allowing the detection of substellar companions around evolved stars through their accretion disks, and potentially even non-accreting hot substellar companions around nearby RGB stars.

The most substantial advance offered by ALMA2040 will be the possibility of studying the effects of planetary engulfment. The theoretical expectation for the evolution of planetary systems is that the large sizes of stars in the giant branches would cause planets in close orbits \citep[within $\sim 2$~au][]{Villaver2014} to be engulfed by the star. In fact,
the dynamical effect of a potentially engulfed planet
has been reported for the nearest AGB star, R~Doradus at $\sim 50$~pc \citep{Vlemmings2018}, 
suggesting an engulfed body with 10 times the mass of Jupiter orbiting at 5~au. Studies of rotation in nearby evolved stars have pioneered this type of observation, but an empirical overview is still lacking as the current ALMA facility only reaches a handful of the closest AGB stars.



\paragraph{Chemical enrichment by evolved binaries:}
Evolved stars are major contributors to the chemical and dust budget of galaxies. For example, AGB stars are the main producers of certain elements (e.g., C and Mg), isotopes (e.g.,  $^7$Li, $^{14}$N, $^{19}$F), and are important contributors of dust grains that get deposited into the ISM \citep[e.g.][]{schneider2024, osborn2025}, and might contribute to dust production already in the early universe, as early as 30--50\,Myr, as recently highlighted in a JWST study \citep{boyer2025_SexA}.
However, models suggest that binary interactions may affect mass-loss rates and suppress or delay key nucleosynthetic yields \citep{osborn2025} and observations show that gas and dust chemistry are impacted by companion presence \citep{Danilovich2024,Danilovich2025}. Quantifying these effects is essential to connect stellar and stellar-system physics to galactic evolution and
sub-millimeter observations are essential to constrain how binary interactions affect mass-loss and nucleosynthetic yields.

\section{Science Case for the 2040s}
\label{sec:sciencecase}

The strength of interactions between evolved stars and their companions increases sharply with decreasing Roche-lobe filling factor. At the widest separations, companions are revealed by their imprint on the outflow; at several to a few tens of au, mass transfer and tides enhance mass loss from the giant primary and alter its evolution over hundreds of thousands of years \citep{Danilovich2025,Decin2021}; and at only a few au, systems enter regimes of dramatic interaction: common-envelope (CE) evolution with rapid ($\sim1$~yr) envelope ejection, mergers, and strongly enhanced accretion rates that produce powerful jets \citep[e.g.][]{Soker2017,Vetter2025}. Planets in the closest orbits are engulfed altogether. The three panels of Fig.~\ref{fig:examples} illustrate these regimes, from AFGL~3068 (centre) through $\pi^1$~Gru (right) to W43~A (left). While current studies focus mainly on post-interaction remnants, the short-lived CE phase itself is expected to become accessible in the 2040s through upcoming time-domain facilities, and real-time observations on 1--10~au scales will enable studies of angular-momentum redistribution, merger events, and the formation and evolution of circumbinary and circumstellar disks.

Several key questions will remain at the forefront of stellar-interaction physics in the 2040s. How do (sub)stellar companions modify wind acceleration and morphology, and the elemental and dust yields of their host stars? How does the evolved primary in turn shape the companion's evolution? Which physical processes govern mass transfer and angular-momentum exchange in evolved binaries? How do the initial separation, eccentricity, and mass ratio affect the outcome? Can second-generation planets form in post-AGB circumbinary disks or around merger remnants?

The large body of knowledge on stellar multiplicity and on the properties of exoplanetary systems allows us to estimate the impact of ALMA2040. Below we follow the sequence of decreasing separation outlined above, giving for each regime an estimate of what would become observable.

\subsection{Detecting the imprints of companions}

The presence of companions of AGB stars with separations up to hundreds of au can currently be detected through high-energy accretion signs for compact companions, such as in D-type symbiotic systems, or for a broader range of companion properties with ALMA through spirals, arcs, and disks \citep[e.g.][]{Maercker2012,kim2017_afgl3068,Decin2020}.
An additional signature currently being pioneered is the impact of companions on the chemical composition of the outflow.

The ultraviolet fields of hotter stellar companions and even X-ray from accreting compact companions initiate photochemistry in the dense inner regions of the outflow, affecting the chemistry throughout the entire circumstellar envelope. Chemical models have identified molecular tracers of binarity for different density structures within the outflow \citep{VandeSande2022,VandeSande2023,Alonso-Hernandez2025}. These have already been successfully used to interpret observations, confirming the presence and type of companion and constraining its orbital parameters \citep{Danilovich2024,Danilovich2025a}. Some of the key molecular tracers are SiN, SiC, and NS.
The study of this binary-induced chemistry requires observation of molecular lines over a broad range of wavelengths, so that excitation and column densities can be constrained. Based on previous detections of these molecules towards three stars \citep{Danilovich2024,Decin2018,Coenegrachts2023,Danilovich2025a}, we find that current ALMA capabilities allow emission to be detected in stars within a few hundred parsecs. The proposed sensitivity of ALMA2040 would extend this to a volume of a few kiloparsecs in radius, encompassing hundreds of systems and dramatically increasing the number of sources for which companion-induced chemistry can be studied.

The nitrogen-bearing molecules have also been detected in the outflows of more massive evolved stars \citep[$\eta$~Car and IRC+10420,][]{Bordiu2022,Quintana-Lacaci2016}, but the achievable angular resolution sets the limit to detailed studies given the large distances, $\gtrsim 2$~kpc. ALMA2040 will provide the necessary higher angular resolution coupled with improved sensitivity.

\subsection{Detecting companions via the accretion disks}
\label{sec:accretion}

For closer companions, with separations up to tens of au, mass transfer becomes the key process to understand. For the very few nearby systems, ALMA2040 would reveal the kinematics of the interaction on sub-au scales, allowing detailed studies of poorly understood processes. As an example, the proposed 3~mas resolution of ALMA2040 would enable mapping of wind--companion interactions for the closest system $\pi^1$~Gru at $0.5$~au resolution and image the cavity of the circumbinary disk of the post-AGB star IRAS~$08544-4431$ with a few resolution elements.

Beyond detailed studies of individual nearby systems, ALMA2040 would give access to whole samples of systems actively exchanging mass, by targeting at (sub-)millimeter wavelengths the most conspicuous signal of an accreting main-sequence companion: dust emission from its accretion disk.
ALMA2040 would be uniquely positioned to detect this emission, revealing the location and properties of companions that at present can only be inferred. By 2040, we expect a large number of candidate systems to have been found through the analysis of upcoming data releases of Gaia.



To estimate the observable effect of an accreting  companion, we consider giants with mass-loss rates of $\gtrsim 10^{-7}$~M$_\odot$~yr$^{-1}$
and assume a conservative WRLOF accretion rate of $5 \times 10^{-9}$~M$_\odot$~yr$^{-1}$.
We consider a separation of 10~au between the two stars and an $\alpha$-disk model \citep{Shakura1973}, with the outer radius set to 30\% of the Roche radius of the companion. The gas-to-dust ratio in the disk is set to 200. In reality, the radiation field of the AGB star will largely determine the disk size and properties, with radiation pressure lowering and dust heating raising the disk brightness; treating these processes is beyond the scope of the present estimates. To determine the number of AGB stars losing mass at a sufficiently high rate within a given volume, we used the volume-limited sample from the NESS survey \citep{Scicluna2022} and selected AGB stars with gas mass-loss rates larger than about $10^{-7}$~M$_\odot$~yr$^{-1}$, finding $\sim 180$ such stars within 300~pc. This number was extrapolated to larger distances considering the scale height of the AGB population to be 300~pc \citep{Jackson2002}.
Assuming a continuum sensitivity at 340~GHz for a 1-hour integration using ALMA2040 of 1~$\mu$Jy, we find that the number of accreting companions that can be imaged through the associated accretion disks would increase by a factor of 50 with ALMA2040 with respect to the current ALMA.

\begin{figure}
{\includegraphics[width=16cm]{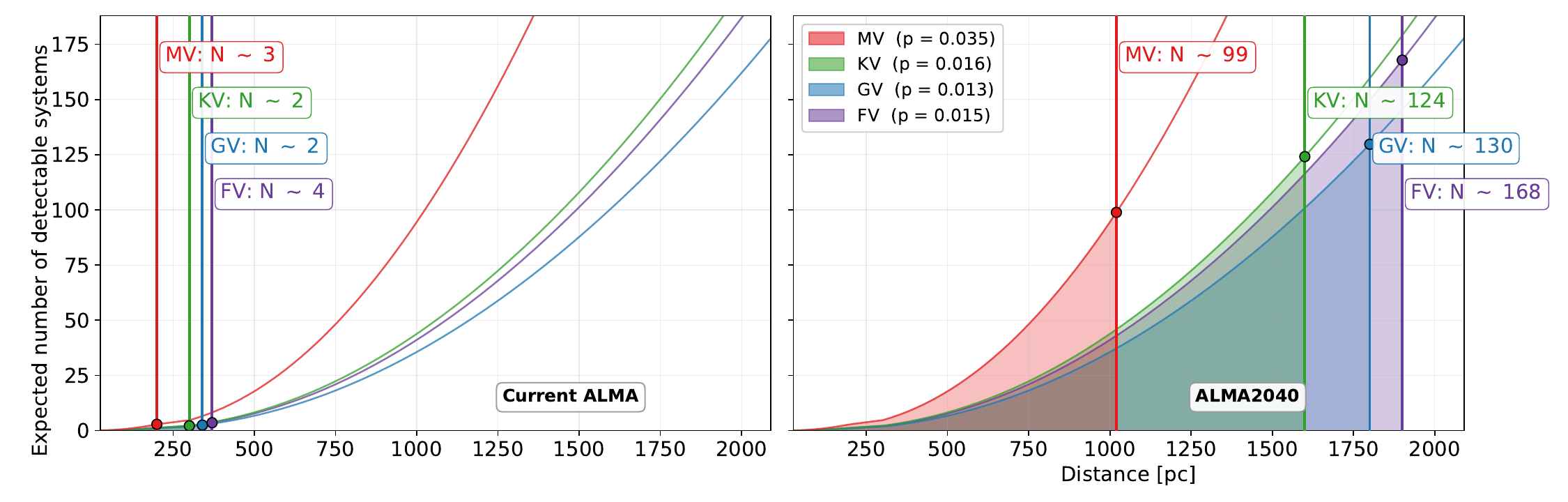}}
\caption{Cumulative distribution of observable systems with different types of accreting main sequence
companions (M, K, G, and F type stars) that can be potentially detected at 340~GHz with 1 hour integration time using ALMA2040 (1~$\mu$Jy sensitivity) and the current ALMA (25~$\mu$Jy sensitivity). ALMA2040 would increase the number of systems studied by a factor of 50, dramatically advancing our empirical knowledge of accretion physics and orbital evolution.}
\label{fig:cumulativeCompanions}
\end{figure}

To estimate the number of systems with different companion types that could be studied, we adopt a typical initial mass of 2~M$_\odot$ for the AGB star and calculate the fraction of primaries with companions of spectral types M, K, G, and F at orbital radii between 5 and 20~au, using the distribution of separation and mass ratios \citep{Moe2017}. We find probabilities of 2.4\%, 1.3\%, 1.4\%, and 2.1\% for a 2~M$_\odot$ star having a companion of spectral type M, K, G, and F, respectively, in the required orbital range.
As shown in Fig.~\ref{fig:cumulativeCompanions}, hundreds of systems will be accessible to ALMA2040, ushering in a new era in the study of binary interactions.
Since these same processes are at play in several areas of astrophysics, this will have far reaching consequences.

\subsection{Catching mergers and engulfment events in real time}
\label{sec:transients}

ZTF~SLRN-2020, the first compelling case of a giant planet merging with a nearly main-sequence star, produced a short optical outburst followed by a much longer-lived infrared excess \citep{De2023} tracing newly formed circumstellar dust \citep{Yarza2025}. The small ejecta and dust masses of planet--main-sequence mergers make their millimetre counterparts challenging to detect, but the prospects improve markedly when the star is evolved and extended: the lower binding energy and cooler ejecta favour higher mass-loss rates and efficient dust formation. Models predict that 30--50\% of coalescence events involving extended, low-mass stars appear as dusty, infrared-luminous transients \citep{MacLeod2022}, and the dust-rich remnants of the Galactic mergers V1309~Sco, V4332~Sgr, and OGLE-2002-BLG-360 provide empirical examples of this regime \citep{Kami2018}. Scaling the observed fluxes of merged systems, a continuum sensitivity of $\sim 1$~$\mu$Jy at $\sim 340$~GHz would reach a Jupiter-mass planet being consumed by a giant host out to $\sim 5$~kpc. Combined with flexible rapid-response target-of-opportunity access, ALMA2040 could follow ongoing low-mass stellar mergers and engulfment events involving subgiants and giants from the onset of the outburst until a few years afterwards, directly tracing the formation, cooling, chemistry, and geometry of the freshly ejected material, and should catch several events in which an evolved star engulfs a giant planet within a decade of operations.

\subsection{Detecting planetary engulfment}

\begin{figure}
\includegraphics[width=16cm]{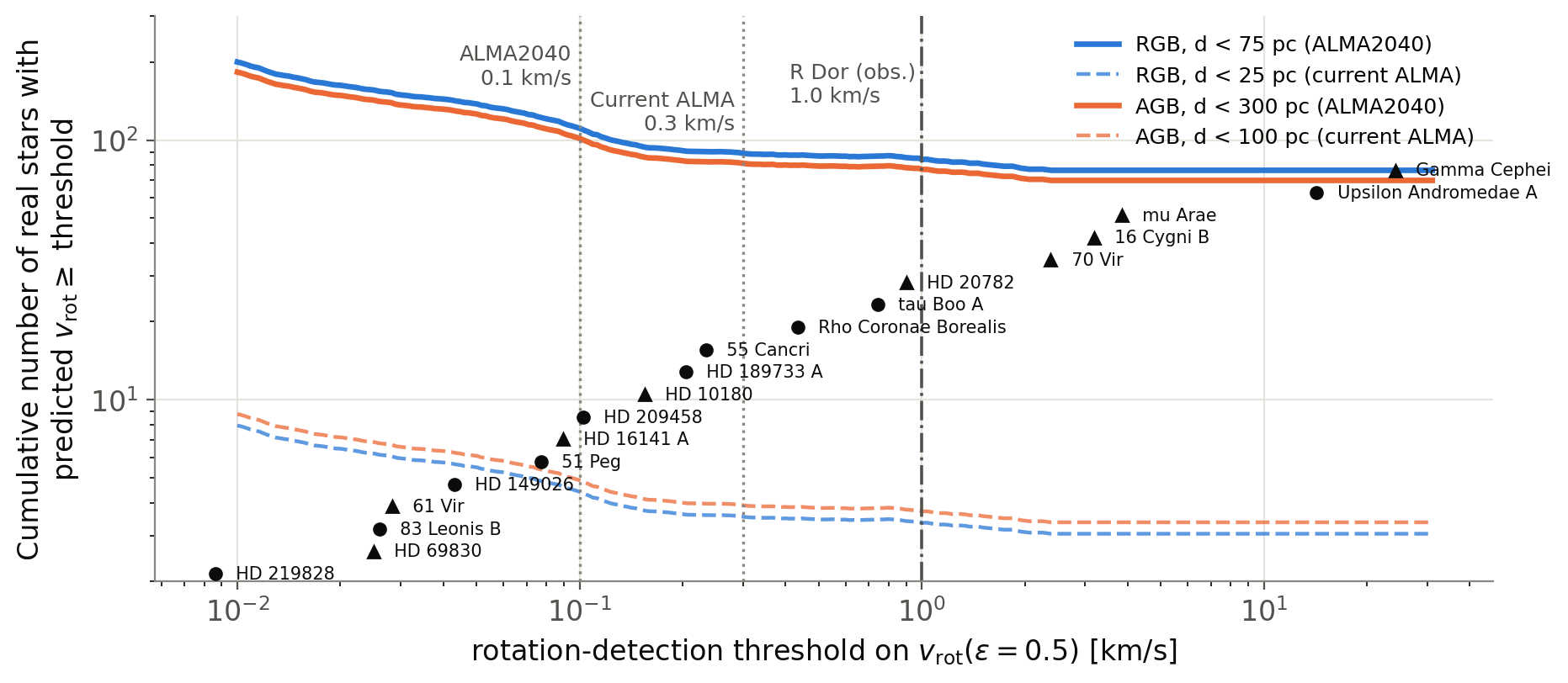}
\caption{Estimated cumulative distribution of systems where planets are expected to have been engulfed by their star. The dashed lines show the distributions for RGB (blue) and AGB (red) stars for current ALMA, while the solid lines show the corresponding curves for ALMA2040. Well-known planetary systems are shown with their respective orbital angular momentum value. $\varepsilon$~is the efficiency of angular momentum transfer and the rotation-detection threshold limits of 0.1 and 0.3~km/s correspond to what will be achieved with ALMA2040 and what is currently possible with ALMA.}
\label{fig:cumulativePlanets}
\end{figure}

The aftermath of planetary engulfment by an expanding star is that the giant is left rotating far faster than it otherwise would \citep{Soker2001}. The excess is orders of magnitude larger than the rotation surviving from early stellar life, which is reduced both by angular-momentum loss and by envelope expansion \citep[e.g.][]{Schrijver1993,Soker2001}. Detecting the excess rotation signature can be done by spatially resolved observations of the stellar disk that reveal rotation through the distribution of velocities in the profiles of spectral lines at (sub-)millimeter wavelengths \citep[e.g.][]{Vlemmings2018,Kervella2018}.
This effect is not only observable but most likely relatively common with planets in orbits of up to a few au expected to be engulfed \citep[e.g.][]{Villaver2014}.
Hence, a large fraction of stars evolving through the giant branches is expected to engulf planets.
Fig.~\ref{fig:cumulativePlanets} shows that ALMA2040 would allow us to study the imprint of engulfed planets in roughly 100 nearby RGB stars and an additional 100~AGB stars. This is a dramatic increase over the few systems for which such an effect can be measured with the current ALMA.

The estimates were based on the mass and separation of exoplanet available on the Open Exoplanet Catalogue \citep{Rein2012}. Then, we constructed a density distribution of planet properties using the determined fraction of exoplanets more massive than one Jupiter mass of $\sim 6.2\%$ \citep{Fernandes2019}. That was used to calculate the probability distribution of stars with different initial masses to acquire a given amount of angular momentum from engulfing planets during the RGB or AGB evolution, based on the maxima radii reached in these two phases. We consider an angular momentum transfer efficiency from the orbit to the stellar envelope of 50\%. Finally, we consider the number of RGB and AGB stars whose stellar disks would be resolvable with ALMA and ALMA2040 for best-possible angular resolutions of 10~mas and 3~mas, respectively.

The rotation of mergers products between evolved and main sequence stars will also be detectable in this way, with significantly higher rotation rates. Tidal spin up by a surviving companion produces a comparable rotation excess to the one estimated for planetary engulfment. In these cases, the companion will be detectable through the accretion signature described in Section~\ref{sec:accretion}.

\vspace{0.2cm}
In conclusion, ALMA2040 will allow the transition from the few pioneering studies of individual cases to a broad overview of statistical samples of evolved stars with companions. This will be fundamental not only to advance the theory explaining these interactions and their consequences to evolved stars, but also for our understanding of the ultimate fate of binary systems and planetary systems like our own.

\section{ALMA2040: Technical Requirements}

The science cases require tracing molecular- and
recombination-line emission and dust and free-free continuum emission at high angular resolution. Typical distances to systems containing AGB (from a few hundred parsecs) and post-AGB (from a few kiloparsecs) stars imply required spatial resolutions of $\sim 3$~mas to resolve the relevant scales discussed above. This is best achieved with a next-generation sub-millimeter interferometer with the following technical requirements: at least $3\times$ better spatial resolution than current ALMA with baselines increased by the same factor, corresponding to an associated $9\times$ smaller PSF area.
A corresponding increase in sensitivity to take advantage of the better resolution, with an increase in instantaneous spectral line sensitivity in Bands 1--10 (35--950\,GHz) of {$10\times$} compared to current ALMA ({$8\times$} compared to anticipated ALMA-WSU). A {$10\times$} increase in continuum sensitivity in Bands 1--10 (35--950\,GHz), compared to current ALMA ({$5\times$} compared to anticipated ALMA-WSU). These can be achieved primarily by increasing the collecting area compared to ALMA-WSU and disruptive
developments in receiver technology, while at the same time providing exquisite image fidelity
(through excellent uv-coverage) across a range of angular scales.
Innovative developments in receiver technology such as
simultaneous multi-band capabilities and ultra-wide bandwidths (full band)
will allow for multi-wavelength time-domain imaging spectroscopy of variable and transient processes, characterizing the physical and chemical impact of (sub)stellar companions on mass loss, and, ultimately, the detection of accreting stellar companions out to $\sim 2$~kpc.

Time-domain studies into orbital properties and orbital evolution require multicycle access to continuous dense $uv$-sampling across spatial scales. Additionally, the study of transient phenomena such as red novae and planet-engulfment events (Sect.~\ref{sec:transients}) requires rapid response times for target-of-opportunity observations, reaching the $\sim 1$~$\mu$Jy continuum sensitivity near 345~GHz needed to follow these events across the Galaxy.

\section*{Acknowledgements}
The authors thank the referees (Pierre Cox and Hans Van Winckel) for their insightful feedback.
This work was developed as part of the broader ALMA2040 community effort. We acknowledge the Lorentz Center and NOVA for supporting the community workshops and coordination activities that helped initiate and shape the ALMA2040 White Paper process.
TK, WV, LP, and BBA acknowledge support from
the Olle Engkvist Foundation through grant 229-0368.
R.S.’s contribution to the research described here was carried out at the Jet Propulsion Laboratory, California Institute of Technology, under a contract with NASA (80NM0018D0004).\\

\noindent
\bibliographystyle{apj_mod}
{
\setlength{\bibsep}{-0.05cm}
\bibliography{references.bib}
}

\end{document}